\documentclass{emulateapj}

\newcommand{\HII}{{\ion{H}{2}}}

\newcommand{\OII}{[{\ion{O}{2}}]}

\newcommand{\OIIIHb}{[{\ion{O}{3}}]/H$\beta$}
\def\ratioR23{([\ion{O}{2}]~$\lambda$3727 +[\ion{O}{3}]~$\lambda\lambda$4959,5007)/H$\beta$}
\def\R23{${\rm R}_{23}$}

\newcommand{\Msun}{${\rm M}_{\odot}$}

\newcommand{\NII}{[{\ion{N}{2}}]}

\newcommand{\OH}{$\log({\rm O/H})+12$}
\newcommand{\NIISII}{[\ion{N}{2}]/[\ion{S}{2}]}

\newcommand{\NIIHa}{[\ion{N}{2}]/H$\alpha$}
\newcommand{\SIIHa}{[\ion{S}{2}]/H$\alpha$}

\newcommand{\SII}{[{\ion{S}{2}}]}

\newcommand{\Hb}{{H$\beta$}}
\def\O4363{[{\ion{O}{3}}]~$\lambda$4363}
\newcommand{\OIII}{[{\ion{O}{3}}]}

\newcommand{\Ha}{{H$\alpha$}}

\def\L60{L$_{60}$}

\newcommand{\lam}{$\lambda$}

\shorttitle{}
\shortauthors{}

\begin{document}

\title{Z-FIRE: ISM properties of the $z = 2.095$ COSMOS Cluster}

\author{Lisa J. Kewley}
\affil{The Australian National University}

\email {lisa.kewley@anu.edu.au}

\author{Tiantian Yuan}
\affil{The Australian National University}

\author{Themiya Nanayakkara}
\affil{Swinburne University}

\author{Glenn G. Kacprzak}
\affil{Swinburne University}

\author{Kim-Vy H. Tran}
\affil{Texas A\&M University}

\author{Karl Glazebrook}
\affil{Swinburne University}

\author{Lee Spitler}
\affil{Macquarie University}
\affil{Australian Astronomical Observatory}

\author{Michael Cowley}
\affil{Macquarie University}
\affil{Australian Astronomical Observatory}

\author{Michael Dopita}
\affil{The Australian National University}
\affil{King Abdulaziz University}

\author{Caroline Straatman}
\affil{Leiden University}

\author{Ivo Labb\'{e}}
\affil{Leiden University}

\author{Adam Tomczak}
\affil{Texas A\&M University}

\begin{abstract}

We investigate the ISM properties of 13 star-forming galaxies within the $z\sim 2$ COSMOS cluster.  We show that the cluster members have \NIIHa\ and \OIIIHb\ emission-line ratios similar to $z\sim 2$ field galaxies, yet systematically different emission-line ratios (by $\sim 0.17$~dex) from the majority of local star-forming galaxies.  We find no statistically significant difference in the \NIIHa\ and \OIIIHb\ line ratios or ISM pressures among the $z\sim 2$ cluster galaxies and field galaxies at the same redshift.  We show that our cluster galaxies have significantly larger ionization parameters (by up to an order of magnitude) than local star-forming galaxies.  We hypothesize that these high ionization parameters may be associated with large specific star formation rates (i.e. a large star formation rate per unit stellar mass).   If this hypothesis is correct, then this relationship would have important implications for the geometry and/or the mass of stars contained within individual star clusters as a function of redshift.

\end{abstract}

\keywords{galaxies:starburst---galaxies:abundances---galaxies:fundamental parameters}

\section{Introduction}\label{Introduction}

Galaxy clusters formed by the collapse of the highest density peaks in primordial density fluctuations \citep[e.g.,][]{Peebles93,Peacock99}.  
The cluster environment strongly affects the star-formation history of cluster members.  Galaxies in the centers of clusters are observed to be redder and more evolved than galaxies on the outskirts of the cluster, as well as field galaxies at the same redshift \citep{Osterbrock60,Dressler85,Balogh97,Balogh99}.  
Thus, the most massive clusters today likely formed in short episodes of vigorous star formation that effectively ceased by $z\sim 1$ \citep{Stanford98,Blakeslee06,vanDokkum07,Eisenhardt08,Mei09,Mancone10,Tanaka13,Brodwin13}.  Theoretical simulations of cluster formation and evolution support this scenario \citep{Springel05,DeLucia06}.   
However, using old stellar populations to reconstruct the star formation conditions and to understand why star formation ceased in clusters is extremely difficult.  Recently, several galaxy clusters have been discovered in the ``redshift desert'' ($1.5<z<2.5$) \citep{Kurk09,Papovich10,Tanaka10,Fassbender11,Gobat11,Nastasi11,Santos11,Spitler12,Stanford12,Zeimann12,Muzzin13,Tanaka13}.  
With near-infrared multi-object spectroscopy on 8-10m telescopes, it is now possible to study the star formation conditions in these clusters, prior to (or during) 
the end of their major star-forming episodes.

Near infrared spectroscopy probes the optical rest-frame emission-lines at $z\sim 2$, providing a wealth of information about the ionized gas surrounding the young, hot stars.  The strong lines \NII, \Ha, \OIII, \Hb, and the \OII\ and \SII\ doublets diagnose the power source of galaxies, the star formation rate, the gas-phase chemical abundance, the ionization state of the gas, the electron density of the gas, and the amount of dust extinction, among other physical properties \citep[see][for an review]{Osterbrock89}.  Knowledge of these properties in clusters across cosmic time helps to build a more comprehensive picture of how star formation progressed in clusters, and how the cluster environment impacted the star formation process.  

So far, large near infrared spectroscopic studies have predominantly focused on field galaxies.  High redshift field galaxies appear to have larger \NIIHa\ and/or \OIIIHb\ optical line ratios (by $\sim 0.3$~dex) compared with local field galaxies \citep{Hainline09,Bian10,Rigby11,Yabe12,Erb06,Zahid13b,Shapley14,Holden14,Steidel14}.  These elevated line ratios may indicate a major change in the ionized gas or star formation properties as a function of time.  Many possible causes for this offset have been proposed, including a larger ionization parameter \citep[e.g.,][]{Brinchmann08b,Shirazi13,Kewley13b}, a harder ionizing radiation field \citep{Steidel14,Stanway14}, a higher ISM pressure (density) \citep{Shirazi13}, a contribution from an AGN \citep{Groves06,Wright10,Trump11}, and contamination by shocks from galactic outflows \citep{Forster13,Newman14}.  Sample selection is likely to play a large role in the relative importance of each of these factors \citep{Juneau14}.

If the cluster environment causes more rapid evolution during the star-formation stage than in the field, then we might expect a difference in the ISM conditions in clusters at high redshift compared with field galaxies.  The star-formation properties of high redshift cluster galaxies relative to the field are currently under debate.  Using stacked near-infrared spectroscopy, \citet{Valentino14} suggested that the relationship between star formation and stellar mass is similar among cluster and field galaxies at $z\sim 2$.  However, \citet{Tran10} and \citet{Brodwin13} argue that the star formation rate in cluster galaxies is elevated relative to field galaxies at the same stellar mass.

 In this paper, we test the hypothesis that star-forming conditions in clusters are more evolved than the field using the $z\sim 2$ COSMOS cluster \citep{Spitler12}.  The COSMOS cluster is an ideal laboratory for studying the formation of massive nearby galaxy clusters.  The COSMOS cluster contains four distinct overdensities \citep{Spitler12}.  Cosmological simulations suggest that at least two of the overdensities will merge across cosmic time, creating a Virgo-like cluster by $z\sim 0$ \citep{Yuan14}.   The cluster has 57 spectroscopically confirmed members, and follows the same mass-metallicity relation as field galaxies at $z\sim2$ (Kacprzak et al. 2014, in prep).  

The COSMOS cluster and comparison samples are described in Section~\ref{Sample}.  In Section~\ref{MOSFIRE}, we describe our near-infrared spectroscopy.  In Section~\ref{BPT_section}, we compare the optical line ratios for 10 galaxies in the COSMOS cluster with field galaxies at the same redshift, and with local star-forming galaxies.   We investigate the electron density (or ISM pressure) in the COSMOS cluster in Section~\ref{Pressure}, and the ionization state of the gas in Section~\ref{Ionization}.  We discuss the implications of this work and our conclusions in Section~\ref{Conclusions}.  Throughout this paper, we adopt the flat $\Lambda$-dominated cosmology as measured by the 7 year WMAP experiment \citep[$h=0.72$, $\Omega_{m}=0.29$;][]{Komatsu11}.

\section{Sample Selection}\label{Sample}

\subsection{The $z\sim 2$ COSMOS Cluster}
The COSMOS cluster was discovered as part of the Magellan/FourStar Galaxy Evolution Survey (ZFOURGE) photometric redshift survey of the COSMOS field \citep{Scoville07a}.  The ZFOURGE survey uses five medium bandwidth filters ($J_{1}$,$J_{2}$,$J_{3}$,$H_{s}$,$H_{l}$) and the broadband $K_{s}$ filter on the Magellan FOURSTAR instrument (Persson13, Straatman et al. in prep).  High-redshift galaxy overdensities were identified using surface density maps within narrow $\Delta z \sim 0.2$ redshift ranges between $1.5<z<3.5$.   With this technique, the COSMOS cluster was revealed with photometric redshifts $z\sim 2.2$ \citep{Spitler12}.  Subsequent Keck observations revealed 57 spectroscopically confirmed cluster members within a projected radius of $\sim 5$Mpc  at $z=2.095$ \citep{Yuan14}.  We compare the ISM properties of the COSMOS cluster with two comparison field samples.

\subsection{Local Comparison Field Sample}

We use the Sloan Digital Sky Survey (SDSS) as our local comparison sample \citep{Adelman06}.  Although the COSMOS cluster galaxies will have evolved into red elliptical galaxies by $z=0$, the SDSS represents the range of properties seen in massive ($M>10^9$\Msun) nearby star-forming galaxies.  We select galaxies from the SDSS DR4, as described in \citet{Kewley06a}.  To avoid aperture effects and incompleteness, we restrict the SDSS redshift range to $0.04<z<0.1$.  The lower redshift limit corresponds to a minimum aperture covering fraction of $\gtrsim 20$\% to minimize aperture effects \citep{Kewley05}.  We use the emission-line fluxes from the JHU/MPA catalog with \Hb, \OIII, \Ha, \NII, and \SII\ emission-lines with signal-to-noise ratios (S/N) $>3\sigma$.  These spectra have been corrected by the JHU/MPA team for underlying stellar absorption using stellar population synthesis models.  Our SDSS sample contains 85,224 galaxies.  We remove AGN using the \citet{Kewley06a} optical classification scheme, leaving 66,590 star-forming galaxies.

\subsection{$z\sim 2$ Comparison Field Sample}

We compare the ISM properties of the COSMOS cluster with the properties of 10 serendipitous field galaxies between $1.8<z<2.2$, observed in the same slitmasks as the COSMOS cluster candidates.   These galaxies were identified in the ZFOURGE five medium bandwidth filters.   

We supplement our small field sample with 39 star-forming galaxies between $1.8<z<2.3$ from the Keck Baryonic Structure Survey (KBSS-MOSFIRE) survey by \citet{Steidel14}.  The KBSS-MOSFIRE star-forming galaxy sample contains $\sim 70.9$\% of galaxies previously selected using rest-frame far-UV spectra, with the remaining galaxies covering a broader range of rest-frame UV and optical colors.  The final KBSS sample with detected \NIIHa\ and \OIIIHb\ emission-lines is not biased in star formation rate and stellar mass, relative to the parent (photometric) sample \citep[see Figure 4 in][]{Steidel14}. 

To avoid selection effects based on galaxy mass, while ensuring statistically significant samples for comparison, we create two comparison field samples, depending on the suite of emission-lines available.  We refer to these two samples as our BPT and \SII\ samples, respectively.  

{\it BPT Field Comparison Sample:} To compare the position of galaxies on the standard optical diagnostic diagrams \citep{Baldwin81,Veilleux87,Kewley01a}, we select galaxies with the full suite of \Hb~$\lambda 4861$, \OIII~$\lambda 5007$, \Ha~$\lambda 6563$, and \NII~$\lambda 6585$ lines.  Steidel et al. apply a lower S/N limit of $5\sigma$ to the \OIII\ and \Ha\ emission-lines, and a $2\sigma$ limit to the \NII\ and \Ha\ emission-lines.   For consistency, we apply the same S/N limits to our field galaxies, but we highlight galaxies with \NII\ or \Ha\ emission-lines with S/N$<3\sigma$ in our Figures to indicate that these detections are marginal.

The position of a galaxy on the optical diagnostic diagrams is sensitive to its gas-phase metallicity, and hence to its stellar mass through the mass-metallicity relation \citep{Tremonti04,Zahid13}.  Selecting a high redshift sample for the full suite of \Hb, \OIII, \Ha, \NII\ emission-lines biases a sample towards galaxies with detectable \OIII\ and \NII\ lines.  This selection can introduce a bias against high mass and low mass galaxies.  Metal-rich nebulae produce low \OIII\ emission-line fluxes and the oxygen lines are sensitive to the electron temperature of the gas, which is cool in metal-rich nebulae.  On the other hand, selecting for \NII\ detections biases against low metallicity galaxies which have weak \NII.  The \NII\ line is sensitive to metallicity only in the metal-rich regime \citep[\OH $\gtrsim 8.4$;][]{Kewley02} because nitrogen changes from a primary to a secondary nucleosynthetic element at intermediate metallicities.  Therefore, we match our field and comparison samples in stellar mass.  Here, we use a subsample of the KBSS-MOSFIRE galaxies matched in stellar mass to our ZFIRE cluster sample with $9.0<{\rm M}_*/$\Msun$<10.0$.  This stellar mass selection results in 17 galaxies.  

Our ZFIRE serendipitious field sample contains 4 galaxies with the full suite of emission-lines \Hb~$\lambda 4861$, \OIII~$\lambda 5007$, \Ha~$\lambda 6563$, and \NII~$\lambda 6585$ detected to signal-to-noise ratios $>3\sigma$ within the same stellar mass range of $9.0<{\rm M}_*/$\Msun$<10.0$.  Applying the KBSS-MOSFIRE S/N limits yields one additional galaxy.

{\it \SII\ Field Comparison Sample:}  We use the \SII\ doublet for calculating the electron density of the ionized gas in the \SII\ zone.  To compare the electron density of the cluster and field galaxies, we 
select galaxies with rest-frame red emission-lines \Ha~$\lambda 6563$, \NII~$\lambda 6585$,  \SII~$\lambda 6717$, and \SII~$\lambda 6731$ detected to the 3$\sigma$ level.   This selection results in high luminosity, high stellar mass samples because the \SII\ lines are only weakly dependent on the metallicity \citep[see e.g.,][]{Kewley02}.  We therefore match our cluster and field galaxies in stellar mass 
($9.5<{\rm M}_*/$\Msun$<11.0$), resulting in 10 field comparison galaxies from our MOSFIRE dataset.   Our BPT and \SII\ field samples are not mutually exclusive, and they span the same range of SFRs ($6<\frac{\rm SFR}{{\rm M}_{\odot}{\rm yr}^{-1}}< 56$).

For our electron density analysis, we are unable to compare with the KBSS-MOSFIRE sample because the KBSS-MOSFIRE \SII\ line fluxes are not publicly available.

\section{ZFIRE: High-z MOSFIRE Near-Infrared Spectroscopy}\label{MOSFIRE}

\subsection{Observations}

We obtained near-infrared spectroscopy for the COSMOS cluster as part of our ZFIRE spectroscopic survey of high-z cluster candidates.  The candidates were selected based on the ZFOURGE photometric redshifts \citep{Spitler12} derived from deep images obtained through the five ZFOURGE near-infrared medium-band filters.  The median uncertainty for the ZFOURGE photometry is $\sim 0.05$~dex \citep{Tomczak14}.  

Observations of cluster candidates were made on Keck MOSFIRE \citep{McLean12} on December 24-25, 2013 and February 10-13, 2014.  Eight masks were used in the $K$-band (1.93-2.45~$\mu$m) to obtain the \Ha\ and \NII\ emission-lines, as well as two masks in the $H$-band (1.46-1.81$\mu$m), to detect the \Hb\ and \OIII\ emission-lines.  Total on-source exposure times were 2 hours for the K-band masks, and 5.3 and 3.2 hours for the two H-band masks.   Seeing varied between 0.4-0.7 arcseconds throughout the exposures.  We used a slit width of 0.7", which gives a spectral resolution of $R=3690$ in the K-band and $R=3620$ in the H-band.  The MOSFIRE field of view of 6.1$\times$6.1 arcminutes enabled 224 objects to be targeted in 6 pointings.   We observed an A0V type standard star in both the wide-slit mode and the narrow-science-slit ($\sim 0. 7$" slit width) mode before and after our science target exposures for telluric and flux calibration.  

\subsection{Data Reduction}

The raw MOSFIRE data were reduced using the publicly-available data reduction pipeline (DRP) developed by the instrument team\footnote{See
\href{http://code.google.com/p/mosfire/}{http://code.google.com/p/mosfire/}}. The DRP provides background-subtracted, rectified and wavelength calibrated 2D
spectra.  All spectra were calibrated to wavelength in a vacuum.  The H-band data were calibrated using night sky lines.  The K-band data were calibrated using both night sky lines and a Neon 
arc lamp to ensure sufficient red emission-lines for wavelength calibration.   The typical residual for the wavelength solution is $\leq 1$\AA.  We used our own custom IDL routines to correct the 2D spectra for telluric absorption, and for flux calibration using 
A0V standard stars.  We have visually inspected the 2-dimensional spectra to check for contamination by sky lines at the positions of the emission-lines, including the weak \NII\lam6584 line and the \SII\lam\lam6716,6731 doublet.  We have also checked that our error spectra fully account for skyline residuals.  The uncertainty for our flux calibration is $\sim$ 8\% (Nanayakkara et al. in prep).  The 1-D spectra and associated 1$\sigma$ error spectra were extracted using an aperture that corresponds to the FWHM of the spatial profile. For objects that are too faint to fit a Gaussian to the spatial profile, we used the FWHM of the stellar profile on the same mask as the extraction aperture. 

\subsection{Emission-Line Measurements}

Gaussian profiles were fit to the emission lines in H and K bands separately.  For widely separated lines such as \OII\lam3727, \Hb\lam4861, single Gaussian functions are fit with 4 free parameters: the centroid (or the redshift), the line width,  the line flux, and the continuum. The doublet \OIII\lam\lam4959,5007 are initially fit as a double Gaussian function with 6 free parameters: the centroids 1 and 2, line widths 1 and 2, fluxes 1 and 2, and the continuum.  A triple-Gaussian function is simultaneously fit to the three adjacent emission lines:  \NII\lam\lam6548, 6583 and \Ha. The centroid and velocity width of the \NII\lam\lam 6548, 6583 lines are constrained by the velocity width of \Ha\lam6563, and  the ratio of \NII\lam6548 and \NII\lam6583 is constrained  to the theoretical value of 1/3 given in \citet{Osterbrock89}.  The line profile fitting is conducted using a $\chi^2$ minimization procedure which uses the inverse of  the 1$\sigma$ error spectrum of the DRP.  This fitting procedure yields the redshift, line flux, line width, continuum and the associated statistical errors.   Our typical 3$\sigma$ flux limit is $1.8\times10^{-18}$~ergs/s/cm$^2$.  Example spectra of our COSMOS cluster members are given in Figure 1 of \citet{Yuan14}. 

From our spectra, we obtained spectroscopic redshifts for 180 galaxies.  Of these, 57 galaxies are confirmed cluster members \citep{Yuan14}.   A total of 8/57 ZFIRE cluster galaxies have the full suite of \Hb~$\lambda 4861$, \OIII~$\lambda 5007$, \Ha~$\lambda 6563$, and \NII~$\lambda 6585$ lines detected to signal-to-noise ratios $>3\sigma$, which we refer to as our BPT cluster sample.   For consistency with the KBSS-MOSFIRE field sample, we also show the positions of galaxies with \Hb~$\lambda 4861$ and \NII~$\lambda 6585$ lines detected to signal-to-noise ratios of $2\sigma$, and \OIII~$\lambda 5007$, \Ha~$\lambda 6563$ lines detected to 
$5\sigma$.  The KBSS-MOSFIRE S/N selection criteria would add 5 additional galaxies to the cluster sample.  We have verified that our results are unchanged regardless of whether we apply our S/N $>3$ criterion, or whether we apply the KBSS-MOSFIRE S/N criteria.

Our ZFIRE BPT cluster sample was selected to satisfy $9.0<\log({\rm M}_*/$\Msun)$<10.0$ to match the mass range spanned by the BPT field control sample.   

Of the galaxies with \Ha~$\lambda 6563$, and \NII~$\lambda 6585$ lines detected to signal-to-noise ratios $>3\sigma$, 4 galaxies also have both \SII~$\lambda 6717$ and \SII~$\lambda 6731$ detected to the 3$\sigma$ level, and span the stellar mass range $9.5<\log({\rm M}_*/$\Msun)$<11.0$, where the stellar mass range is matched to the \SII\ comparison sample.  We refer to these four \SII\ detected cluster galaxies as our \SII\ cluster sample.   Of these, only 3 galaxies have all of the \Hb, \OIII, \Ha, \NII, and \SII, emission-lines detected to the $3\sigma$ level.

\subsection{Stellar Mass and SFR measurements}

We compute stellar masses using the ZFOURGE $J_{1}$,$J_{2}$,$J_{3}$,$H_{s}$,$H_{l}$) and $K_{s}$ photometry.  We fit the \citet{Bruzual03} stellar population synthesis models with FAST \citep{Kriek09}, assuming a \citet{Chabrier03} initial mass function, a \citet{Calzetti00} attenuation law, and solar metallicity.   The same models and parameters were used by \citet{Steidel14} for the KBSS-MOSFIRE sample.   Our BPT cluster sample, the Steidel BPT field sample, and our ZFIRE BPT field sample have the same mean stellar masses, within the errors ($<\log({\rm M}_*/$\Msun$)>= 9.62\pm0.06, 9.63\pm0.08$, and $9.67 \pm 0.07$, respectively).   Although limited by small numbers, our \SII\ cluster and field comparison samples also have consistent stellar masses, within the errors ($10.0\pm0.2$ c.f. $9.6 \pm 0.2$, respectively).

Star formation rates (SFRs) are calculated from the extinction-corrected \Ha\ emission-line and the \citet{Hao11} calibration, as described in detail in Tran et al. (2015, ApJ, submitted).  We correct the \Ha\ emission-line for extinction using the \citet{Cardelli89} extinction curve.
The resulting SFRs range between $5<\frac{\rm SFR}{{\rm M}_{\odot} {\rm yr}^{-1}}<110$ for our BPT cluster sample and $5<\frac{\rm SFR}{{\rm M}_{\odot} {\rm yr}^{-1}}<225$ for our BPT field sample.  The 2D Kolmogorov-Smirnov test indicates that our cluster and field samples have statistically consistent (P(KS)$< 1\sigma$) extinction and star-formation rates, given the sample sizes and errors.  We do not match the star formation rate range for the cluster and field galaxies because the star formation rate is correlated with the ionization parameter, and the ionization parameter is one of the quantities that we aim to test for differences among field and cluster galaxies.

\section{Optical Emission-Line Analysis}\label{BPT_section}

\subsection{The \NIIHa\ versus \OIIIHb\ Diagnostic Diagram}

The \NIIHa\ versus \OIIIHb\ diagram  was originally proposed by \citet{Baldwin81} to distinguish between galaxies powered by \HII\ regions, planetary nebulae, and objects powered by a harder ionizing radiation field, such as galaxies containing active galactic nuclei (AGN).  This diagram (known as the BPT diagram) is useful at intermediate and high redshift because the \NII\, \Ha\, \OIII\, and \Hb\ lines are often the only emission-lines observable in near-infrared spectra of high redshift galaxies, and the wavelength of the emission-lines are sufficiently close that the \NIIHa\ and \OIIIHb\ ratios do not need to be flux calibrated or corrected for extinction. 

\begin{figure}[!t]
\epsscale{1.0}
\plotone{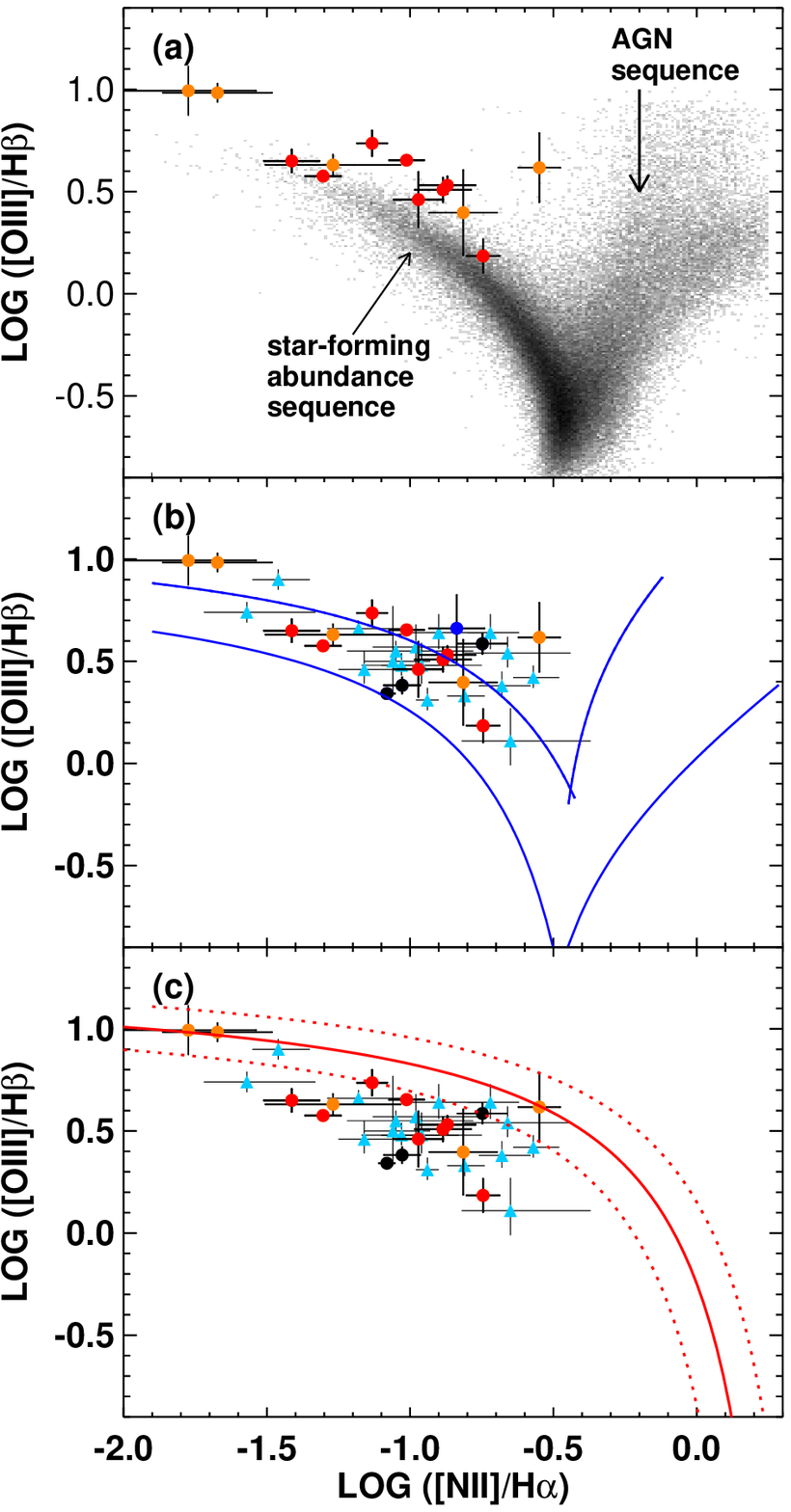}
\caption[f1.eps]{The \NIIHa\ versus \OIIIHb\ diagnostic diagram showing our COSMOS cluster (red circles) in comparison to the local SDSS sample (upper panel), our field comparison sample (middle panel), and the \citet{Kewley13b} redshift-dependent classification line and error ranges (red solid line and dotted lines respectively, lower panel). The blue lines in the middle panel show the position of the local SDSS envelope for comparison.  Field galaxies are colored according to sample, where pale blue corresponds to the KBSS-MOSFIRE field sample of \citet{Steidel14}, and black corresponds to field galaxies from our MOSFIRE observations with all strong-lines detected to $>3\sigma$.   Orange circles and dark blue circles show the additional galaxies that would be added to our cluster sample and field sample if we were to apply the S/N limits of Steidel et al.  
\label{BPT}}
\end{figure}

Both \HII\ regions and star-forming galaxies form a curved locus on the \NIIHa\ versus \OIIIHb\ optical diagnostic diagram \citep{Dopita00,Kewley06a}.  In Figure~\ref{BPT}(a), we show the locus of the star-forming galaxies in the SDSS sample.   \citet{Veilleux87} showed that theoretical photoionization models can reproduce this curved locus.  This locus is now known as the star-forming abundance sequence because the shape of the locus is dictated primarily by the chemical abundance (metallicity) of the star-forming regions within galaxies.  Metal-poor galaxies have low \NIIHa\ and high \OIIIHb\ ratios due to the combination of low chemical abundance and the electron temperature sensitivity of the oxygen collisionally excited lines.   

The ISM and star formation properties of a galaxy affects its position relative to the star-forming abundance sequence on the BPT diagram \citep[see Figure 1 in][]{Kewley13a}.  A rise in the ISM pressure causes both the \NIIHa\ and the \OIIIHb\ line ratios to rise.  A hard ionizing radiation field (such as from a Wolf-Rayet dominated stellar population) causes a similar rise in the \NIIHa\ and \OIIIHb\ line ratios.

However, a rise in the ionization parameter causes the \NIIHa\ (and \SIIHa) to fall, while \OIIIHb\ rises.  The ionization parameter is defined as the number of hydrogen ionizing photons passing through a unit area per second divided by the number density of hydrogen atoms $n_H$, and is a measure of the amount of ionization that a radiation field is able to produce in an \HII\ region.
For a spherical geometry, the ionization parameter $q$ can be defined using the Stromgren radius $R_s$ \citep{Stromgren39}:

\begin{equation}
q =\frac{Q_{{\rm H}^{0}}}{4 \pi {{R}_s}^2 n_H}  \label{q_spherical}
\end{equation}

where $Q_{{\rm H}^{0}}$ is the flux of ionizing photons above the Lyman limit.  In an ionized nebula, the number density of hydrogen is approximately the electron density $n_e$.  

The ionization parameter is affected by both the hardness of the ionizing radiation field, as well as the bolometric luminosity of the ionizing source.  The combination of theoretical photoionization models, and observations of the \OIII, \Hb, \NII, \Ha\ lines, and the \SII\ doublet allows one to discriminate between the effect of a differing ionization parameter, the ISM pressure, and the radiation hardness.  The \SII~$\lambda6717/$\SII~$\lambda6731$ ratio constrains the ISM pressure, thanks to the sensitivity of this doublet to the hydrogen density of the gas.  The \SIIHa\ ratio is particularly sensitive to the hardness of the ionizing radiation field because the \SII\ lines are produced in a partially ionized zone that is large when the radiation field contains a significant fraction of high energy (X-ray and EUV) photons.  Finally, the combination of the \NIISII\ and \OIIIHb\ ratios separate the ionization parameter and the metallicity, as shown in \citet{Dopita13}.

Figure~\ref{BPT}(a) gives the \NIIHa\ versus \OIIIHb\ diagram for the SDSS (grey contours) and the COSMOS cluster (red symbols).  The SDSS star-forming sequence is parameterized by \citet{Kewley13a} (their equation 5).  Using this parameterization, we calculate the median \OIIIHb\ offset between the COSMOS cluster and the SDSS star-forming sequence to be $\sim 0.17$~dex.  We investigate possible causes for this offset in the following sections.

We compare the position of our COSMOS cluster with our field comparison sample in Figure~\ref{BPT}(b).  We have matched the stellar mass range of both samples (i.e. $9.0<\log({\rm M})/{\rm M}_{\odot} < 10.0$) to avoid mass biases.  

Our COSMOS cluster has line ratios that are well matched to the field galaxies at the same redshift. The two-sided Kolmogorov-Smirnov statistic for the \NIIHa\ and \OIIIHb\ ratios yields D-values of 0.34 and 0.35, respectively, with significance levels of 0.36 and 0.32, indicating that our cluster and the field sample are not significantly different.   We have also verified that the relationships between \NIIHa\ and stellar mass, and \OIIIHb\ and stellar mass are the same for the cluster and field samples, within the errors.  We conclude that the cluster environment has not produced more rapid evolution in the ISM conditions in our star-forming galaxies at $z\sim 2$.

\subsection{AGN and shock contamination}

AGN strongly affect the \NIIHa\ and \OIIIHb\ optical line ratios.  Galaxies containing AGN form a separate sequence on the BPT diagram, beginning at the high metallicity, high mass end of the star-forming abundance sequence (Figure~\ref{BPT}a).  This AGN sequence occurs because the hard radiation field from an AGN produces more ionizations into the ${\rm O}^{++}$ ion, and more collisional excitations of the \OIII\ line.   The position of an AGN host galaxy along the AGN sequence depends on the contribution from star-formation to the line ratios, the slope of the EUV radiation field, and the ionization parameter of the radiation field \citep{Groves04b}.  

We use the redshift-dependent theoretical optical classification scheme of \citet{Kewley13b} to identify possible AGN in our sample
(Figure~\ref{BPT}c).  This scheme was based on the predictions of cosmological hydrodynamic simulations, as well as stellar evolutionary synthesis and photoionization models.  We find that three galaxies in our cluster with $2<$S/N(\NII)$<3$ may have a small ($\sim 20$\%) contribution from an AGN based on their optical line ratios (i.e. \OIIIHb$\gtrsim 1.0$), but these galaxies are within the $\pm 0.1$~dex error range of the classification line (dotted lines in Figure~\ref{BPT_NIIHa}c), and may be low metallicity star-forming galaxies with a hard ionizing radiation field, such as from Wolf-Rayet stars \citep{Kewley01a,Dopita06b,Masters14}, or their \NII\ line fluxes may simply have too low S/N to reliably classify.  We discuss these galaxies further in Section~\ref{Ionization}.  The remaining galaxies in our cluster are likely to be purely star-forming.

\begin{figure}[!t]
\epsscale{0.8}
\plotone{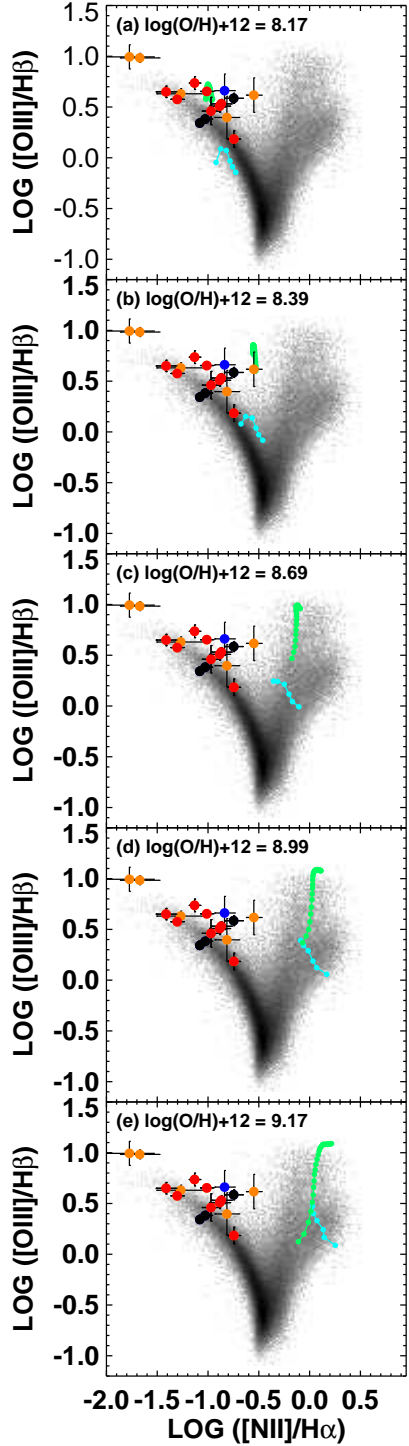}
\caption[f2.eps]{Allen et al. (2008) fast shock models (green) and the Rich et al. (2010) slow shock models (cyan) as a function of metallicity (panels a-e).  The shock model data indicate where galaxies that are 100\% dominated by shocks are likely to lie.  The SDSS sample from \citet{Kewley06a} is shown in grey.  The positions of  COSMOS cluster galaxies (red circles) and COSMOS field galaxies (black circles) are not consistent with the 100\% shock models for the metallicities spanned by these galaxies (\OH$\gtrsim 8.3$).  Orange and dark blue circles show the additional galaxies that would be added to our cluster sample and field sample if we were to apply the S/N limits of Steidel et al.  
\label{Shocks_BPT}}
\end{figure}

Shocks can also produce strong \OIIIHb, and can mimic AGN on the BPT diagram locally  \citep[e.g.,][]{Kewley01b,Rich11}.  At the low metallicities of high redshift galaxies, shocks can also mimic the emission-line ratios of star-forming galaxies \citep{Kewley13a}.  In Figure~\ref{Shocks_BPT}, we show the position of our COSMOS cluster galaxies relative to the slow shock models of \citet{Rich10,Rich11} (cyan) and the fast shock models of \citet{Allen08} (green).  These models are described in detail in \citet{Kewley13a}, and examples of typical mixing sequences are given in Rich et al. (2010, 2011). The shock models occupy different regions of the diagnostic diagram depending on the metallicity of the sample.  Only the lowest metallicity (\OH$<8.17$) shock models can account for the line ratios of our cluster galaxies.  However, our cluster galaxies lie along the standard galaxy mass-metallicity relation at $z\sim 2$ \citep{Kacprzak15}, with metallicities between $8.3<$\OH$<8.7$ in the \citet{Kewley02} metallicity scale.  Here, we have converted the metallicities from the \citet{Pettini04} calibration into the \citet{Kewley02} scale using the calibrations provided in \citet{Kewley08} to yield consistent metallicity estimates with our photoionization and shock models.  Because the shock models for $8.3<$\OH$<8.7$ produce optical line ratios that are significantly larger than observed (by 0.1-0.4 dex), we rule out a dominant contribution from shocks to the optical emission-line ratios of our cluster galaxies.

Galaxies containing a mixture of ionizing sources, such as shocks and star formation will lie along mixing sequences between the star-forming sequence and the 100\% shock models. The position and shape of the mixing sequences depend on the metallicity of the galaxy and the shock velocity. 
We cannot rule out a small ($\sim$10\%) contribution from shocks to the emission-lines ratios.  High angular resolution and high spectral resolution integral field spectroscopy with adaptive optics can distinguish shocked regions from star-forming regions at high redshift \citep{Yuan12}.  This integral field spectroscopy is one of our future research directions for this sample.

\subsection{Comparison with theoretical models of ISM conditions}\label{models}

We compare the location of our ZFIRE cluster on the BPT diagram with our predictions from theoretical stellar population synthesis and photoionization models.  Our models are described in detail in \citet{Kewley13a} and \citet{Dopita13}.   We use the chemical evolution predictions from \citet{Dave11b} to constrain the gas-phase metallicity history for star-forming galaxies with $M_* >10^{9}$~\Msun\ .  The metallicity at a given redshift determines the shape of the EUV spectrum produced in the Starburst99 or Pegase2 evolutionary synthesis models \citep{Leitherer99,Fioc99}, which in turn, affects the line intensities predicted by the Mappings IV photoionization code \citep{Dopita13}.  We use the instantaneous and continuous burst models from Starburst99 and Pegase2 with a Salpeter Initial Mass Function, and the Pauldrach/Hillier stellar atmosphere models.   We note that the choice of IMF makes negligible difference in the line ratios that we are using in this analysis \citep{Dopita13b}.   The resulting stellar population spectra are embedded within a spherical nebula.  Our Mappings IV photoionization code calculates radiative transfer self-consistently throughout the nebula.  Mappings IV uses a Kappa temperature distribution which is more suitable for a turbulent ISM than a Stefan-Boltzmann distribution \citep{Nicholls12}.  For AGN, we use the dusty radiation-pressure dominated models of \citet{Groves04a}.  

With these models, we produce theoretical predictions for how the optical emission-line ratios will appear for galaxy samples at different redshifts, based on the following limiting assumptions for star-forming galaxies and AGN:

\begin{itemize}

\item Star forming galaxies at high redshift may have ISM conditions and/or an ionizing radiation field that are either (a) the same as local galaxies ({\it normal ISM conditions}) or (b) more extreme than local galaxies ({\it extreme ISM conditions}).  Extreme conditions in star-forming galaxies can be produced by a larger ionization parameter and a denser interstellar medium, and/or an ionizing radiation field that contains a larger fraction of photons able to ionize O$^{+}$ into O$^{++}$ (i.e. energy $>35.12$~eV) relative to the number of Hydrogen ionizing photons (i.e. a harder radiation field).  In \citet{Kewley13a}, we use an ionizing radiation field from the Pegase 2 stellar population synthesis models \citep{Fioc99} to mimic the hard ionizing radiation field  from the stellar atmospheres from massive stars with the effects of stellar rotation.   The difference between our hard ionizing radiation field and our normal ionizing radiation field varies with wavelength \citep[see Figure 2 in][]{Kewley01a}, between $1-10 ergs/s/$\Msun in $\log (\lambda {\rm F}_{\lambda})$ between $100-1000$\AA.  Similar effects on the optical line ratios can be achieved by increasing the ionization parameter by up to an order of magnitude (i.e. $q=2\times 10^{7}$~cm/s to $q=2\times 10^{8}$~cm/s, and/or by raising the electron density by up to two orders of magnitude (i.e. $n_{e} = 10$~cm$^{-3}$ to $n_{e} = 1000$~cm$^{-3}$) (Kewley et al. in prep).

\item The AGN narrow line region at high redshift may either (c) have already reached the level of enrichment observed in local galaxies ({\it metal-rich}), or (d) have the same metallicity as the surrounding star-forming gas.  In the latter case, the AGN narrow-line region at high redshift would be more {\it metal-poor} than local AGN narrow-line regions.  Our AGN narrow-line region models cover a large range of power-law indices ($-1.2<\alpha<-2.0$) and ionization parameters ($0.0<\log(U)<-4.0$).  Therefore, the theoretical AGN region may be larger than the observed location of AGN on the BPT diagram at high redshift.

\end{itemize}

These two sets of limiting assumptions yield four limiting scenarios for where galaxies might be located on the BPT diagram at different redshifts: 

\begin{enumerate}
\item Normal ISM conditions, and metal-rich AGN narrow-line regions at high-z.
\item Normal ISM conditions, and metal-poor AGN narrow-line regions at high-z.
\item Extreme ISM conditions, and metal-rich AGN narrow-line regions at high-z.
\item Extreme ISM conditions, and metal-poor AGN narrow-line regions at high-z.
\end{enumerate}

\begin{figure*}[!t]
\epsscale{1.2}
\plotone{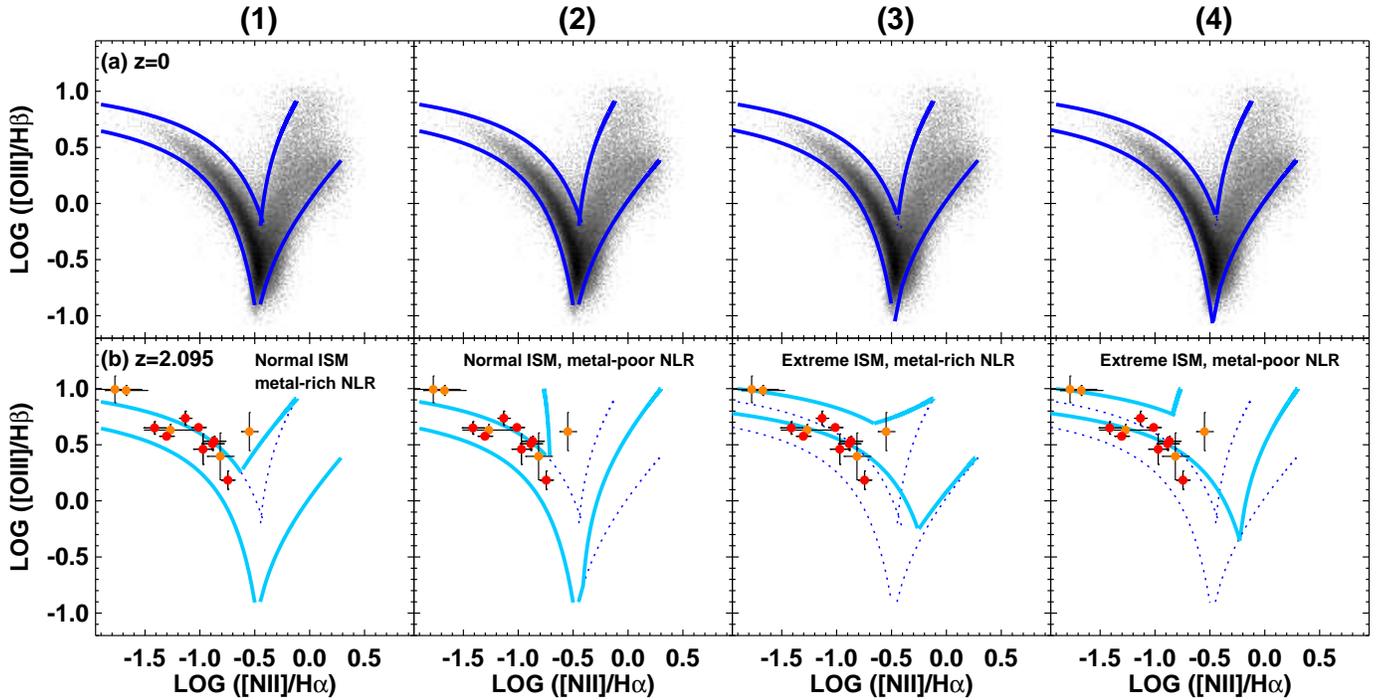}
\caption[f3.eps]{The \NIIHa\ versus \OIIIHb\ diagnostic diagram showing the local SDSS galaxies (top panel), and our $z=2.095$ COSMOS cluster (bottom panel; red circles).  Orange and dark blue circles show the additional galaxies that would be added to our cluster sample and field sample if we were to apply the S/N limits of Steidel et al.  
Solid lines show our theoretical predictions for the star-forming abundance sequence (left curves) and the starburst-AGN mixing sequence (right curves) for z=0 (top) and z=2 (bottom) for 4 limiting model scenarios (columns). Column (1): Normal ISM conditions, and metal-rich AGN NLR at high-z; Column (2): Normal ISM conditions, and metal-poor AGN NLR at high-z; Column (3): Extreme ISM conditions, and metal-rich AGN NLR at high-z; Column (4): Extreme ISM conditions, and metal-poor AGN NLR at high-z.   Blue dotted lines in the lower panel show the position of the local models, for comparison.  
\label{BPT_NIIHa}}
\end{figure*}

 Our theoretical model predictions for each of these four limiting scenarios are given by the solid lines in Figure~\ref{BPT_NIIHa} for $z=0$ (upper panel) and $z\sim 2$ (lower panel).  The COSMOS cluster appears to span a range of ISM and star-forming conditions, from nearly local-star-forming conditions (panels 1, 2) to more extreme star-forming conditions (panels 3, 4).  Our data do not contain sufficient AGN to distinguish between the two limiting AGN scenarios.  We conclude that at least 4 of our cluster galaxies have more extreme star forming conditions than local star-forming galaxies, on average, and that their star-forming conditions are similar to field galaxies at the same redshift, as seen in Figure 1.

\section{Electron Density/Pressure of the ISM}\label{Pressure}

We investigate whether the \NIIHa\ ratio correlates with the electron density of the ionized gas.  The mean ISM pressure, $P$, is related to the total gas density, $n$, and mean electron temperature, $T_e$, through $n =\frac{ P }{ T_e k }$.  Here, the total density $n$ is related to the electron density through $n=2 n_{e} (1+H_{e}/H)$.  In a fully ionized plasma, the electron temperature is $\sim 10^4$~K and the ISM pressure is directly proportional to the electron density \citep[see e.g.,][for a discussion]{Dopita06}.   The mean ISM pressure is related to the mean mechanical luminosity flux from the central star clusters within \HII\ regions.  In the standard adiabatic shell model for \HII\ regions, \HII\ regions expand until the internal pressure from mass-loss and supernova energy-driven bubbles equals the ambient pressure of the ISM \citep{Oey97}.  Therefore, in the expanding bubble model, the ISM pressure is inversely proportional to the \HII\ region radius \citep{Dopita06}.

We calculate the electron density using the \SII\ doublet ratio, \SII~$\lambda 6717$/\SII~$\lambda 6731$.  We assume a four-level model atom in our Mappings~IV models.  The relationship between the \SII~$\lambda 6717$/\SII~$\lambda 6731$ and electron density given by Mappings~IV is identical to the \SII-electron density relationship in \citet{Osterbrock89}.  When the \SII\ ratio approaches $\sim 1.45$, the electron density is in the low density limit (i.e., $\lesssim 10$~$cm^{-3}$). Only one out of five of the COSMOS cluster galaxies with measurable \SII\ ratios is in the low density limit.

\begin{figure}[!t]
\epsscale{1.2}
\plotone{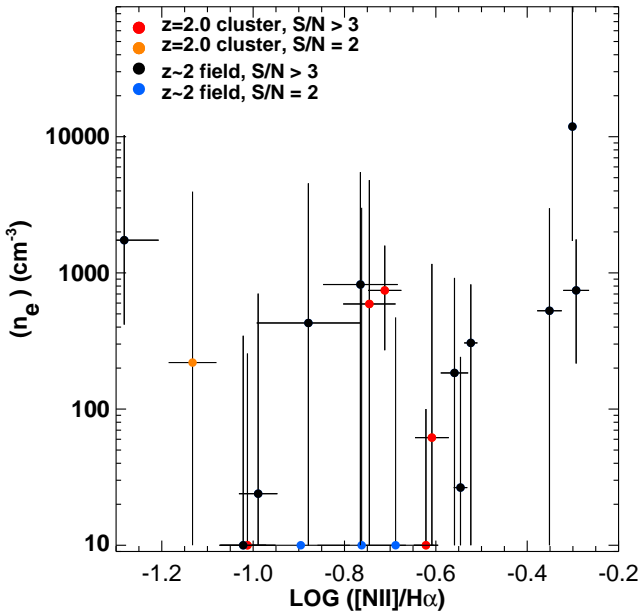}
\caption[f4.eps]{The \NIIHa\ ratio versus electron density for the COSMOS cluster (red circles), and field galaxies 
observed as part of the ZFIRE survey (black circles).  Also shown are additional galaxies that would be added to our cluster sample and field sample if we were to apply the S/N limits of Steidel et al. (orange and dark blue circles, respectively).
\label{NIIHa_ne}}
\end{figure}

In Figure~\ref{NIIHa_ne}, we show the electron density versus the \NIIHa\ ratio for the COSMOS cluster and the $z\sim 2$ field galaxies from our \SII\ sample.  The cluster galaxies span a similar range of electron densities as the field galaxies at the same redshift.  The two-sided Kolmogorov-Smirnov statistic yields a D-value of 0.4, with a significance level of 0.54, indicating that the two samples are not significantly different.   Clearly, there is no systematic difference in the electron density among our cluster and field galaxies at $z\sim 2$, within the limitations of our small sample size and relatively large errors.

\section{The Ionization State of the Gas}\label{Ionization}

The ionization state of the gas (or "ionization parameter") is traditionally measured using the ratio of two ionization states of the same atomic species.  The most commonly used ionization parameter diagnostic is the \OIII~$\lambda 5007$/\OII~$\lambda 3727$ ratio.  This ratio is usually calibrated as a function of ionization parameter and metallicity using stellar population synthesis and photoionization models \citep[e.g.,][]{Kewley02,Kobulnicky04}.  Here, we use the \NIISII\ and \OIIIHb\ ratios to distinguish the ionization parameter from metallicity, because the rest-frame blue \OII\ line is not available within the wavelength coverage of our spectra.   The \NIISII\ ratio is sensitive to ionization parameter because the ionization potentials of the \NII~$\lambda 6584$ and \SII~$\lambda \lambda 6717, 31$ lines differ significantly.  The ratio of \NIISII\ is also sensitive to metallicity, particularly for metallicities below \OH$<8.69$.  This metallicity sensitivity occurs because the ratio of collisional excitation rates of \NII\ and \SII\ is a weak function of nebular temperature, which becomes high at low metallicities due to the lack of coolants in the nebula.    We note that although sulphur is a typical alpha element (i.e. its abundance scales directly proportional to oxygen) \citep{Ryde14}, nitrogen has both primary and secondary nucleosynthetic origins, and is sensitive to the metallicity in the secondary nucleosynthetic regime (\OH$\gtrsim 8.4$).  To account for this effect, our models use an empirical function for N/O versus O/H which takes into account both the primary and secondary nucleosynthetic components, as described in \citet{Dopita13}.  The \OIIIHb\ ratio helps to separate the ionization parameter from metallicity, particularly below \OH$<8.69$, where the \NIISII\ ratio becomes insensitive to ionization parameter.  In this regime, the \OIIIHb\ ratio is primarily sensitive to the ionization parameter.  At higher metallicities, \OH$>8.69$, the oxygen lines in the optical and the infrared dominate the cooling of the nebula, leading to less collisional excitations of the ${\rm O}^{++}$ ion, and a stronger dependency between \OIIIHb\ and metallicity.   

\begin{figure}[!t]
\epsscale{1.2}
\plotone{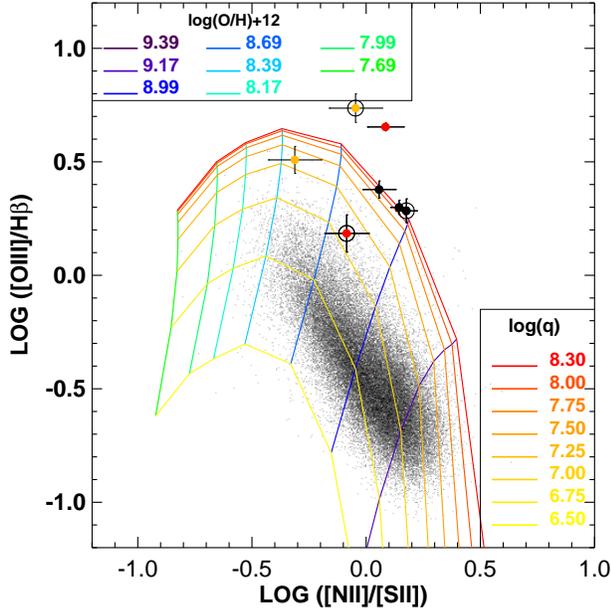}
\caption[f5.eps]{The \NIISII\ ratio versus \OIIIHb\ for the $z\sim 2$ COSMOS cluster for $>3\sigma$ detections of [SII] (red circles) and for $\sim 2\sigma$ detections of [SII] (orange circles), compared with our $z\sim2$ ZFIRE field galaxies (black filled circles) and the local SDSS sample (dots).  The stellar mass range of the SDSS sample has been matched to the stellar mass range of our $z\sim 2$ galaxies 
($9.5<\log({\rm M})/{\rm M}_{\odot}<11.0$).  The colored curves show our theoretical photoionization models for a range of metallicity (blue-green lines), and ionization parameter (yellow-red lines), as indicated in the legends.  The COSMOS cluster and field galaxies have larger ionization parameters than the majority of local SDSS galaxies.  Galaxies with electron densities that could be as large as $n_e \sim 1000$~cm$^{-3}$ within their errors are highlighted with a bold circular outline.
\label{NIISII_OIIIHb}}
\end{figure}

In Figure~\ref{NIISII_OIIIHb}, we show theoretically how the ionization parameter (yellow-red curves) and the metallicity (green-blue curves) are related on the \NIISII\ versus \OIIIHb\ diagram. For ionization parameters $\log(q) < 7.75$, the \NIISII\ and \OIIIHb\ line ratios successfully separate the ionization parameter from metallicity.   We show cluster and $z\sim 2$ field galaxies with $>3\sigma$ detections of [SII] (red circles and black circles respectively), as well as cluster galaxies with less reliable $2\sigma$ detections of [SII] (orange circles).  The majority of local star-forming galaxies have ionization parameters between $6.0< \log(q)< 7.25$, while our COSMOS and field galaxies have larger ionization parameters, i.e. $\log(q) \geq 7.25$.  The two galaxies with the largest \OIIIHb\ ratios lie above the theoretical curves, suggesting a possible small (i.e. $\sim 20$\%) AGN contribution.  Additional information from the X-rays, IR, or radio is required to confirm the presence of an AGN in these galaxies.

The electron density of our models \citep[described in][]{Dopita13b} is fixed by the ISM pressure, and corresponds to $n_{e} \sim 10$~cm$^{-3}$.  An electron density of  $10$~cm$^{-3}$ is consistent with all of our cluster galaxies and 2/3 of our field galaxies shown on Figure~\ref{NIISII_OIIIHb}, within the errors.  We note that if the model electron density is increased from $10$~cm$^{-3}$ to an extreme value of $1000$~cm$^{-3}$, the theoretical \OIIIHb\ ratio would rise by $\sim 0.4$~dex.   The \SII\ line fluxes are subject to large uncertainties (up to 1000~${\rm cm}^{-3}$).  Within these large errors, two cluster galaxies and one field galaxy could potentially have electron densities as large as $1000$~cm$^{-3}$.  We highlight the position of these three galaxies on Figure~\ref{NIISII_OIIIHb} with a circular outline.  Even with an $1000$~cm$^{-3}$, two out of the three high electron density galaxies would still have ionization parameters larger than local SDSS galaxies at the same stellar mass.

Large ionization parameters  have been seen in high redshift galaxies in previous work \citep[e.g.,][]{Brinchmann08b,Shirazi13,Steidel14},  but the cause of such a large ionization parameter is unknown.  \citet{Brinchmann08b} and \citet{Shirazi13} suggest that higher electron densities and a larger escape fraction of H-ionizing photons may be responsible for the larger ionization parameter.  On the other hand, \citet{Steidel14} suggest that a high effective temperature in the massive stellar population, possibly from stellar rotation and/or binaries may be responsible for the large ionization parameters.  Our sample is too small to reach statistically significant conclusions on the cause of the high ionization parameter in our cluster and field galaxies, especially given that four of our galaxies lie along the edge of the ionization parameter-metallicity grid in Figure~\ref{NIISII_OIIIHb} where distinguishing between ionization parameters above $\log(q) > 7.75$ is impossible.  We note that the cluster galaxy with the lowest ionization parameter ($\log(q) \sim 7.4$) has a significantly lower specific star formation rate ($\log ({\rm SFR / M_*}) \sim -8.9$) than the majority of our cluster galaxies (mean $<\log ({\rm SFR / M_*})> = -8.4\pm 0.2$).  If there is a relationship between specific star formation rate and ionization parameter, then this may have important implications on the geometry and mass of the star-forming clusters at high redshift.  

\section{Conclusions}\label{Conclusions}

In this paper, we have investigated the rest-frame optical line ratios of the COSMOS galaxy cluster at redshift $2.095$.  We compare the line ratios of the COSMOS cluster with field galaxies at the same redshift.  We incorporate theoretical stellar evolution, photoionization, and shock models into our analysis to investigate any changes in ISM conditions, ISM pressure, or ionization parameter as a function of redshift or as a function of environment.  

We find that the COSMOS cluster optical line ratios are indistinguishable from field galaxies at the same redshift.   We find no statistical difference among the ISM pressure or ionization parameter of field or cluster galaxies, within the limited size of our sample.  We conclude that for the COSMOS cluster, the cluster environment makes no or limited impact on the ISM properties of its member galaxies.  We also rule out shocks as a dominant contributor to the optical emission-lines in our cluster galaxies.

We show that the COSMOS cluster has larger \NIIHa\ and/or \OIIIHb\ ratios than local galaxies, similar to the large line ratios seen in field galaxies at high redshift.  This offset is consistent with a theoretical change in ISM conditions with redshift.  We investigate possible causes of this change of ISM conditions using our theoretical models.  We rule out a change in ISM pressure as the dominant cause of the extreme ISM conditions in our high redshift sample.  We use the \NIISII\ versus \OIIIHb\ diagram to distinguish the ionization parameter from metallicity.  We show that the large line ratios are likely to result from a large ionization parameter ($\log(q) \geq 7.25$).  The ionization parameters in our COSMOS cluster and field galaxies are significantly larger (by up to 1 dex) than seen in local galaxies.  We suggest that there may be a relationship between ionization parameter and specific star formation rate.  A larger sample of high-redshift galaxies with ionization parameter and specific star formation rate measurements is required to test this idea and to explore possible implications on the \HII\ region geometry and star cluster properties.

\acknowledgments

The authors thank the referee for an excellent and comprehensive referee report.  L.J.K. gratefully acknowledges the support of an ARC Future Fellowship and ARC Discovery Project DP130103925.  L.J.K. and M.A.D. also acknowledge the support of ARC Discovery Project DP130104879.  The writing of this paper would not have been possible without the ANU CHELT 2014 Academic Women's Writing Workshop.  L.J.K. thanks the RSAA interstellar plotters and the GEARS-3D group for providing a supportive and stimulating environment under which to conduct this research.    G.G.K acknowledges the support of the Australian Research Council through the award of a Future Fellowship (FT140100933).  This research has made use of NASA's Astrophysics Data System Bibliographic Services and the NASA/IPAC Extragalactic Database (NED).   Data in this paper were obtained at the W.M. Keck Observatory, made possible by the generous financial support of the W.M. Keck Foundation, and operated as a scientific partnership among Caltech, the University of California and NASA.  We recognize and acknowledge the very significant cultural role and reverence that the summit of Mauna Kea has always had within the indigenous Hawaiian community.  


\end{document}